\documentclass[aps,prl,twocolumn]{revtex4}

\usepackage{graphicx,url}

\begin{document}
\title{Fruit flies modulate passive wing pitching to generate in-flight turns}

\author{Attila J. Bergou$^1$}
\email{ajb78@cornell.edu}
\author{Leif Ristroph$^1$}
\author{John Guckenheimer$^2$}
\author{Itai Cohen$^1$}
\author{Z. Jane Wang$^3$}
\email{zw24@cornell.edu}
\affiliation{$^1$Department of Physics, Cornell University, Ithaca, New York
14853, USA\\
$^2$Department of Mathematics, Cornell University, Ithaca, New York 14853,
USA\\
$^3$Theoretical and Applied Mechanics, Cornell University, Ithaca,
New York 14853, USA}

\date{\today}

\begin{abstract}
    Flying insects execute aerial maneuvers through subtle manipulations of
    their wing motions.  Here, we measure the free flight kinematics of fruit
    flies and determine how they modulate their wing pitching to induce sharp
    turns.  By analyzing the torques these insects exert to pitch their wings,
    we infer that the wing hinge acts as a torsional spring that passively
    resists the wing's tendency to flip in response to aerodynamic and inertial
    forces.  To turn, the insects asymmetrically change the spring rest angles
    to generate rowing motions of their wings.  Thus, insects can generate these
    maneuvers using only a slight active actuation that biases their wing
    motion.
\end{abstract}

\maketitle

\begin{figure*}
    \centering
    \includegraphics{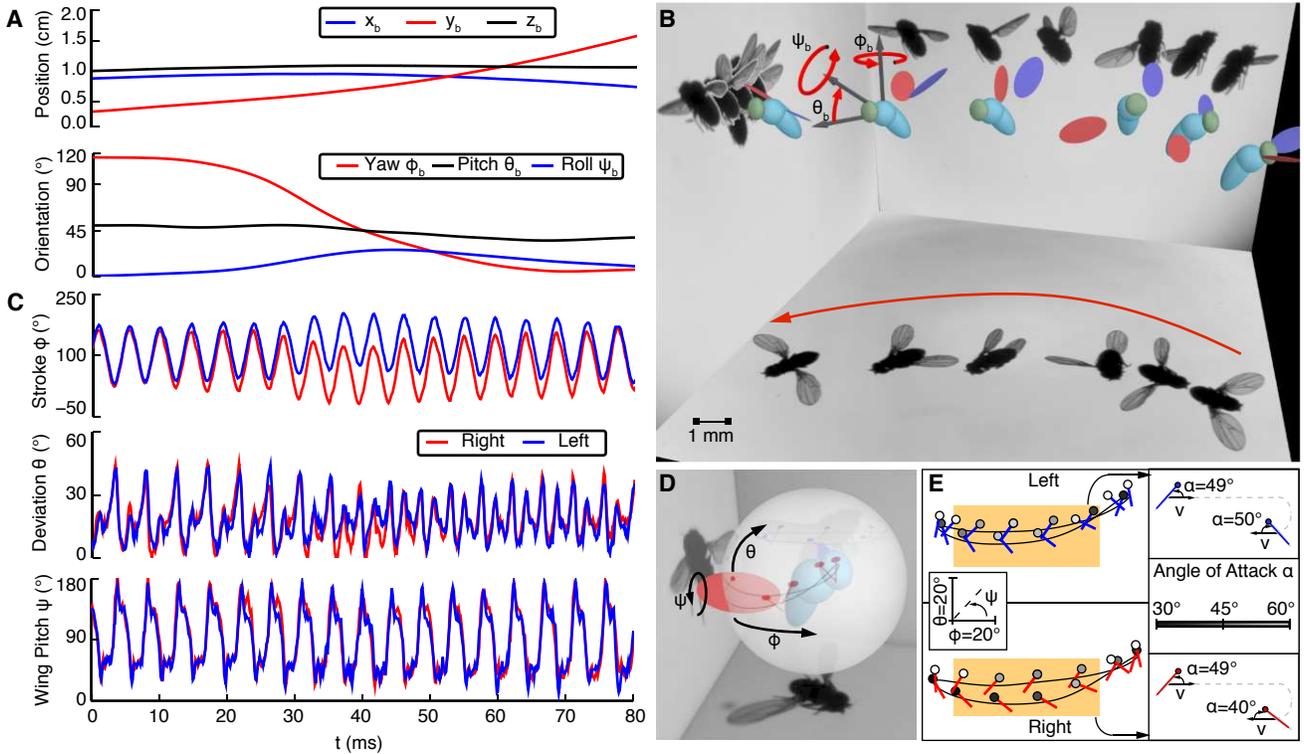}
    \caption{
    ({\bf A}) Measured fruit fly body position and orientation versus time.
    ({\bf B}) Phases of motion of a fruit fly performing a $120^\circ$ turn.
    The three panels show 6 of 821 frames recorded by high speed videography,
    while the corresponding reconstructed kinematics are visualized using the
    model flies.  Labeled on the fifth phase are the Euler angles that quantify
    the fly's orientation.  ({\bf C}) Measured fruit fly wing orientation
    relative to body versus time.  ({\bf D}) The trajectory and orientation of
    wing chords during flapping stroke.  The red and blue lines on the globe
    depict the chord's orientation at equal time snapshots, with the ball
    depicting the wings leading edge.  Labeled on the globe are the Euler angles
    used to quantify wing orientation.  ({\bf E}) The unwrapped ball-and-stick
    diagram highlights the $9^\circ$ asymmetry in the mid-stroke angle of attack
    that causes the maneuver. }
    \label{fig:experiment}
\end{figure*}
To generate the vertical force necessary to sustain flight, small insects must
beat their wings hundreds of times per second. Under this constraint, how do
they manipulate these fast wing motions to induce flight maneuvers? Although
recent studies have made progress addressing how wing motions generate
aerodynamic forces \cite{Fry:2003p917, Ellington:84a, Sane:2003p66,
Wang:2005p289}, understanding how the wing motions themselves arise and what
control variables govern them remains a challenge. To address these questions,
we analyze the torques freely-flying fruit flies ({\em D.\ melanogaster}) exert
to move their wings.  Specifically, we elicit sharp free-flight turns from these
flies and measure their wing and body kinematics.  By using a model of the
aerodynamic forces on flapping wings, we extract the torques the insects exert
to generate the wing motions.  From these torques, we construct a mechanical
model of the wing rotation joints that demonstrates how the interplay of
aerodynamic, inertial and biomechanical forces generate the wing kinematics.
Finally, we connect this model to the turning dynamics of flies and describe the
wing actuation mechanism that unifies these maneuvers.

To quantify turning kinematics in fruit flies, we first use three orthogonal
cameras to capture their free flight maneuvers at 8000 frames per second or
about 35 frames for each wing beat.  We then reconstruct the three-dimensional
wing and body motion of the flies from these videos using the motion tracking
techniques described in \cite{Ristroph:09a}. The body kinematics, described by
the centroid coordinates and three Euler angles - yaw, $\phi_b$, body pitch,
$\theta_b$, and roll, $\psi_b$, are shown in Fig.\ \ref{fig:experiment}A and
visualized in Fig.\ \ref{fig:experiment}B.  During the level flight, the fly
performs a $120^\circ$ turn in 80 ms, or 18 wing beats.

To induce such a turn, the insect generates differences between the motion of
its left and right wings.  We quantify these changes by plotting in Fig.\
\ref{fig:experiment}C the time course of three Euler angles - stroke, $\phi$,
deviation, $\theta$, and wing pitch, $\psi$ - that describe the orientation of
the wings relative to the hinges they rotate about \cite{Wootton:1992p113}. A
three dimensional representation of a typical wing stroke is shown in Fig.\
\ref{fig:experiment}D. Remarkably subtle asymmetries between the pitch angles of
the wings drive the turn.  For a discussion of why the other observed
asymmetries produce a negligible effect on the yaw dynamics see
\cite{Bergou:2009p0}. To quantify the wing pitch asymmetry that induces the
maneuver we phase-average \cite{Revzen:2008p051907} six consecutive strokes from
$t=10$ ms to $30$ ms, during which these asymmetries are most prominent.  
We find that the symmetric front- and back-strokes of the left wing result in
canceling drag forces and no net torque on the fly.  
In contrast, the right wing beats with mid-stroke angles of attack
$\alpha=49^\circ$ during the front- and $\alpha=40^\circ$ during the back-
stroke.  
The $9^\circ$ difference in $\alpha$ results in a net drag force towards the
back of the fly and a torque that turns the fly clockwise.  Thus, by varying
wing pitch and consequently angle of attack, insects row through the air to
perform a turn.

To elucidate how the fly controls its wing pitch, we invert the equations of
motion of the wings to determine the torque the fly exerts at the wing hinge,
\begin{equation}\label{eq:taubase}
  \vec\tau_{i} = \mathbf{I}_w \cdot \dot{\vec\omega} - \vec{r} \times m_w
  \vec{a} - \vec\tau_a.
\end{equation}
From kinematic data, we determine the rotational acceleration $\dot {\vec
\omega}$ and acceleration $\vec a$ of the wing centroid.  We also measure the
wing mass $m_w$ and use morphological measurements of fruit fly wings to
estimate the wing moment of inertia $\mathbf I_w$ and center of mass to hinge
vector $\vec{r}$ \cite{Ennos:1988p199}.  The aerodynamic torque about the wing
hinge, $\vec\tau_a$, is calculated using a quasi-steady model detailed in
\cite{Pesavento:2004p167, Bergou:2009p0}.  From Eq.\ (\ref{eq:taubase}), we
determine the torque component flies exert to pitch their wings, $\tau_p$,
during the turn and during steady flight.

\begin{figure}
    \centering
    \includegraphics{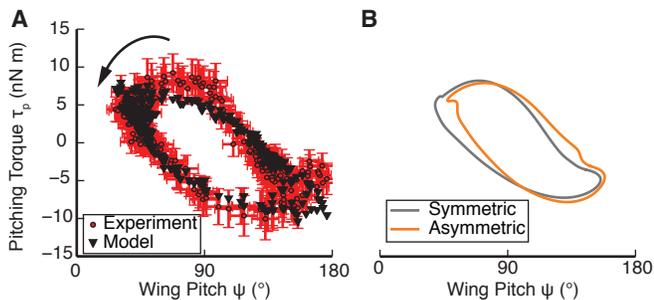}
    \caption{({\bf A}) The pitching torque $\tau_p$ versus the wing pitch
    $\psi$ for the wing beats associated with steady flight.  Red circles
    represent instantaneous values that correspond to frames of the flight
    sequence.  Black triangles are the result of fitting $\tau_p$ with a damped
    torsional spring model.  The counter-clockwise direction of propagation
    indicates that the motion is damped.  ({\bf B}) The phase-averaged $\tau_p$
    versus $\psi$ for strokes associated with symmetric wing movements (gray
    trajectory) compared to asymmetric wing movements (orange trajectory).}
    \label{fig:dynamics}
\end{figure}

Previous studies have suggested that the pitching torques result from torsional
deformation of the wing hinge and simply resist the tendency of aerodynamic and
inertial forces to flip the wing \cite{Ennos:1988p199, Miyan:1988p2386,
Ennos:1988p2424, Bergou:07a, Ishihara:2009p2415, Mountcastle:2009p}. To
integrate these ideas into a dynamical model for the pitching torque, we plot
$\tau_p$ versus $\psi$ in Fig.\ \ref{fig:dynamics}A for the final 9 consecutive
strokes associated with steady flight in Fig.\ \ref{fig:experiment}C.  The
torque data traces out an elliptical curve whose major axis has a negative
slope.  This negative correlation indicates that when the wing angle deviates
from approximately $90^{\circ}$, the hinge produces a restoring torque.  The
area enclosed by the ellipse indicates the energy dissipated by the hinge as it
pitches the wing.  These data suggest the hinge acts like a damped torsional
spring,
\begin{equation}
  \tau_p = -\kappa \left(\psi - \psi_0\right) - C \dot \psi,
  \label{eq:spring}
\end{equation}
where the parameters $\kappa$, $C$, and $\psi_0$ correspond, respectively, to
the torsion constant, damping constant and the rest angle of the torsional
spring.  We fit this model to all 9 wing strokes, finding values $\kappa = 91
\pm 9 \; \mathrm{pN \: m/^\circ}$, $C = 39 \pm 12 \; \mathrm{fN \: m \:
s/^\circ}$, and $\psi_0 = 90 \pm 1^{\circ}$ that account for 95\% in the
variance of the pitching torque (Fig.\ \ref{fig:dynamics}A).

\begin{figure}
    \centering
    \includegraphics{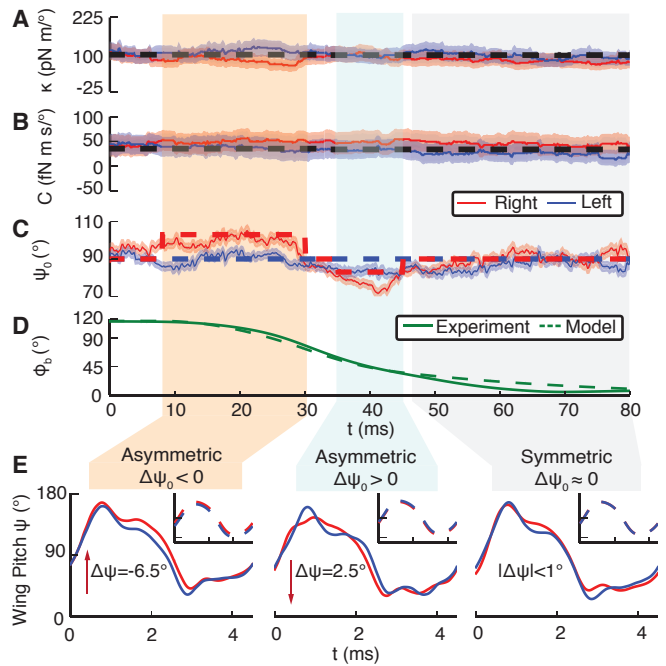}
    \caption{({\bf A})--({\bf C}) Torsional spring parameters versus time
    extracted with error bars from the experiment data. ({\bf D}) Yaw angle
    versus time. The solid-line corresponds to experimental measurements and
    dashed-line to simulation of a model fly that flaps with zero deviation
    angle and a sinusoidal stroke angle with amplitude and frequency that match
    the measured values.  In the simulation, $\kappa$ and $C$ are held constant
    as in (A) and (B) and $\psi_0$ is varied as the dashed lines in (C) ({\bf
    E}) Phase-averaged pitch of the right and left wings, $\psi$, for one
    complete stroke in the three labeled regions.  Solid lines are experimental
    measurements, while dashed lines in the insets are corresponding strokes
    from the model fly.  The shift $\Delta\psi$ in the three regions are
    $6.5^\circ$ (orange $\Delta\psi_0<0$), $2.5^\circ$ (blue $\Delta\psi_0>0$),
    and $<1.0^\circ$ (gray $\left|\Delta\psi_0 \right| < 1^\circ$).  The
    corresponding $\Delta\psi$ for the model fly are $8^\circ$, $2.9^\circ$ and
    $0^\circ$, respectively.}
    \label{fig:ideal}
\end{figure}

To determine how the wing pitch is actuated differently during the turn, we
repeat the above analysis for the 6 strokes that induce the maneuver.  We find
that Eq.\ (\ref{eq:spring}) also accounts for the exerted torques during these
wing beats.  We compare the phase-averaged torque for the 6 asymmetric strokes
with the phase-averaged torque for the 9 symmetric steady strokes in Fig.\
\ref{fig:dynamics}B.  We find that the two loops are shifted horizontally with
respect to each other, indicating a change in $\psi_0$, the wing's rest angle.
In fact, by plotting the values for $\kappa$, $C$ and $\psi_0$ as a function of
time, we show that only $\psi_0$ varies significantly throughout the maneuver
(Fig.\ \ref{fig:ideal}A--C).  The steadiness of $\kappa$ and $C$ suggests that
these parameters correspond to material properties of the wing hinge.
This interpretation is supported by the fact that the value of the spring
constant, $\kappa$, agrees with scaled estimates for the torsion constants taken
from measurements on the wings of house flies \cite{Ennos:1988p2424}.
Comparison of the parameter values with the fly's yaw dynamics in Fig.\
\ref{fig:ideal}D indicates that, to induce the clockwise turning maneuver
(orange shading), the insect increases the $\psi_0$ of the right wing relative
to the left by about $15^{\circ}$ for 6 strokes.  We find that for the two
subsequent wing beats (blue shading) the insect decreases the $\psi_0$ of the
right wing relative to the left by about $10^{\circ}$.  This reversal in the
sign of $\Delta \psi_0 = \psi_0^{(l)} - \psi_0^{(r)}$ indicates that the insect
is generating a counter-clockwise torque that slows its yaw velocity. In the
final 9 wing strokes (gray shading), the $\psi_0$ values for the left and right
wings are nearly equal so that no active torque is generated.

To validate that the yaw dynamics of the fly derives from changes to the
relative rest angle of the wings, $\Delta\psi_0$, we simulate the coupled
wing-body dynamics in a model fly.  The driving of each wing is simulated by
prescribing its stroke and deviation angles.  We then determine the fly's yaw
angle and wing pitch angles from the interaction of the spring model in Eq.\
(\ref{eq:spring}) with the aerodynamic forces on the wings \cite{Bergou:2009p0}.
To form a minimal model of the turn, the variables $\kappa$ and $C$ are held
constant (dashed lines in Figs.\ \ref{fig:ideal}A,B), $\theta$ is set to zero,
and $\phi$ is driven sinusoidally with flapping amplitude and frequency that
match the experiments.  The rest angle, $\psi_0$, for the left and right wings
are prescribed, respectively, by the blue and red dashed lines in Fig.\
\ref{fig:ideal}C, which capture the observed trends.  Further, we ensure
that locally the area between the dashed lines matches the area enclosed between
the two experimental curves.  In Fig.\ \ref{fig:ideal}D we compare the observed
yaw dynamics of the insect to those predicted by our minimal turning model.  We
find that the turn predicted by the model fly (dashed line) quantitatively
matches experimental measurements (solid line) capturing the total accumulated
yaw angle and time-course of the turn.  In Fig.\ \ref{fig:ideal}E, we compare
the measured wing pitch with our model predictions (insets) for the three
regions that correspond to $\Delta\psi_0 < 0$, $\Delta\psi_0 > 0$ and
$\Delta\psi_0 \approx 0$.  We find that the minimal model recovers the average
measured wing pitch asymmetries, $\Delta\psi$, in these three regions, but does
not reproduce the detailed sub-wing beat variations in $\psi$.  When we extend
our simulation to prescribe $\phi$ and $\theta$ of the wings as measured
experimentally, these details are also reproduced \cite{Bergou:2009p0}.  The
fact that the turning dynamics of the fly are predicted by the minimal model and
are insensitive to the details of the stroke and deviation angles of the wing
beat strongly indicates that the turn is controlled by $\Delta\psi_0$.  This
finding is further underscored by the fact that simulation of experimentally
measured $\phi$ and $\theta$ angles without asymmetries in the rest angle,
$\Delta\psi_0 = 0$, does not lead to asymmetric wing pitch angles and therefore
does not yield a turn \cite{Bergou:2009p0}.

\begin{figure}
    \centering
    \includegraphics{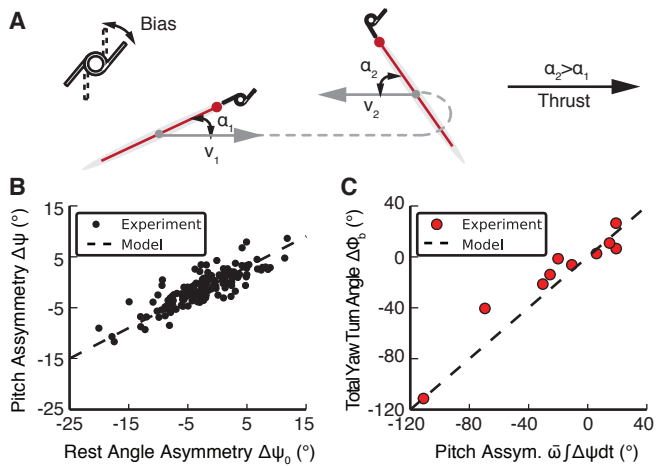}
    \caption{ ({\bf A})  The minimal model of turning dynamics: the wings are
    driven with a sinusoidal stroke, while yaw and wing pitch are determined by
    aerodynamic, inertial, and spring torques.  Biasing the rest angle of the
    spring, produces an asymmetric stroke that generates thrust.  A difference
    in rest angles, $\Delta \psi_0$, of the left and right wings induces a turn.
    ({\bf B}) The average values of $\Delta\psi$ versus $\Delta\psi_0$ for 147
    strokes (black points).  We find $\Delta\psi = \mu \Delta\psi_0$, where $\mu
    = 0.51 \pm 0.03$ from experiment, and $\mu = 0.6$ from simulations of the
    minimal turning model.  Measurement errors are on the order of the spread of
    the data and are omitted for clarity.  ({\bf C}) The total turn angle
    of the fly versus cumulative pitch asymmetry.  Measurements of ten turns
    (red dots) are compared with prediction of the minimal turning model (dashed
    line).  Measurement errors are on the order of the marker size and are
    omitted for clarity.}
    \label{fig:summary}
\end{figure}
Taken together our findings show that as flies flap their wings back and forth,
the wing hinges resist the tendency of aerodynamic and inertial forces to flip
the wings. In steady flight, this passive mechanism causes the pitching of
insect wings and results in front- and back- strokes with nearly symmetric
angles of attack.  The rest angles of the wings act as control levers that break
this symmetry (Fig.\ \ref{fig:summary}A).  This model reconciles how insects
combine active and passive modulation of their wing kinematics to control flight
\cite{Pettigrew:1873, Marey:1895, Ennos:1988p199, Ennos:1988p2424,
Miyan:1988p2386, Bergou:07a}.  Our simulations show that the shift in relative
rest angles produces a linear shift between the wing pitch angles so that
$\Delta\psi \approx \mu \Delta\psi_0$ (Fig.\ \ref{fig:summary}B).  We find from
experimental analysis of 147 distinct wing strokes that $\mu = 0.51 \pm 0.03$
which compares well with simulations of the sinusoidally flapping model fly
where $\mu = 0.6$ \cite{Bergou:2009p0}.  The asymmetry in wing pitch angles
causes asymmetric drag forces on the wings, which are related to the fly yaw
dynamics by:
\begin{equation}
  I_b \ddot\phi_b + 2 \bar\omega C_\tau \dot\phi_b = 2 C_\tau \bar\omega^2
  \Delta\psi,
  \label{eq:dynamics}
\end{equation}
where $I_b$ is the moment of inertia of the fly about the yaw axis, $C_\tau$ is
a parameter that depends on the wing drag coefficient and geometry, $\bar\omega$
is the average angular velocity of the wings, and $\phi_b$ is the yaw of the
body \cite{Bergou:2009p0}.  When $\Delta\psi = 0$, Eq.\ (\ref{eq:dynamics})
reduces to the passively damped yaw motion characteristic of flapping flight
\cite{Hesselberg:2007p2970, Hedrick:2009p2892}.  When $\Delta\psi \neq 0$, the
fly generates a driving torque that actively yaws the fly.  By integrating Eq.\
(\ref{eq:dynamics}) over time, we find the accumulated yaw,
\begin{equation}
  \Delta\phi_b = \bar\omega \int\Delta\psi dt 
  \approx \bar\omega \mu \int\Delta\psi_0 dt.
  \label{eq:summary}
\end{equation}
We confirm that Eq.\ (\ref{eq:dynamics}) accounts for the observed yaw
dynamics of flies by plotting $\Delta\phi_b$ versus $\bar\omega \int\Delta\psi
dt$ for 10 turning maneuvers in Fig.\ \ref{fig:summary}C. We find that the yaw
angles of the turns are in excellent agreement with the prediction of Eq.\
(\ref{eq:summary}).  Thus, by changing the strength and duration of the
asymmetry in wing rest angles, flies can accurately control their turn angle.  

In all forms of locomotion - aquatic, terrestrial, and aerial - animals take
advantage of mechanical properties of their bodies to simplify the complex
actuation necessary to move \cite{Dickinson:2000p100, Liao:2003p1566,
Collins:2005p2953}.  Here, we find a concrete example of how this simplification
occurs as fruit flies modulate their wing motion to generate maneuvers.  The
mechanical properties of the wing hinge appear to be fine-tuned to enable
insects to modulate their wing pitch through only slight active actuation.  The
spring-like behavior of the wing hinge also connects the timescale of a turning
maneuver with the timescale of the wing actuation.  Our model predicts, that
flight muscles of flies can act over several wing beats to bias the pitch of the
wings and yet generate the sub wing-beat changes in wing motion that
aerodynamically induce the maneuver.  Finally, because animals over a wide range
of length scales experience similar rotational dynamics
\cite{Hedrick:2009p2892}, the simple mechanism used by fruit flies may be quite
general and should likewise simplify the control of flapping flying machines.

\begin{acknowledgments}
We thank Andy Ruina, Gordon Berman, Kirk Jensen, Song Chang for helpful
discussions.  This work is supported by NSF.
\end{acknowledgments}

\bibliographystyle{apsrev}

\end{document}